# Cluster Size Optimization in Cooperative Spectrum Sensing

Ebrahim Karami and Amir H. Banihashemi

Department of Systems and Computer Engineering, Carleton University, Ottawa, ON, Canada

ebkarami, ahashemi@sce.carleton.ca

*Abstract—* In this paper, we study and optimize the cooperation cluster size in cooperative spectrum sensing to maximize the throughput of secondary users (SUs). To calculate the effective throughput, we assume each SU spends just 1 symbol to negotiate with the other SUs in its transmission range. This is the minimum overhead required for each SU to broadcast its sensing decision to the other members of the cluster. When the number of SUs is large, the throughput spent for the negotiation is noticeable and therefore increasing the cooperation cluster size does not improve the effective throughput anymore. In this paper, we calculate the effective throughput as a function of the cooperation cluster size, and then we maximize the throughput by finding the optimal cluster size. Various numerical results show that when decisions are combined by the OR-rule, the optimum cooperation cluster size is less than when the AND-rule is used. On the other hand, the optimum cluster size monotonically decreases with the increase in the average SNR of the SUs. Another interesting result is that when the cluster size is optimized the OR-rule always outperforms the AND-rule.

## I. INTRODUCTION

Wireless spectrum is limited and currently most of the available spectrum has been allocated through fixed spectrum allocation. A recent report [1] by USA Federal Communications Commission (FCC) indicates that the current fixed spectrum allocation uses the spectrum very inefficiently and in other words, the spectrum is utilized in a sparse manner. Spectrum surveys from the other countries have also confirmed the FCC report [2]. On the other hand, new wireless services require more spectrum access opportunities. Cognitive radio provides a chance to take advantage of the sparsity of the fixed allocated bands [3, 4]. A cognitive radio is flexible to change its parameters to interact with its environment [5]. In a cognitive radio system, such as the one shown in Fig. 1, users who are licensed to use the spectrum are called *primary users* (PU) and cognitive radios are called *secondary users* (SU).

Spectrum sensing (SS) is the first necessary step to take advantage of the available spectrum [6]. The objective of the spectrum sensing is to determine whether the spectrum is used by the PUs or it is free. A few sensing methods have been proposed for the spectrum sensing such as energy detection [6-9], classic maximum likelihood ratio test (LRT) [9], matched filtering [7, 9, 10], cyclostationary detection [11-13], and others [14-16]. Spectrum sensing based on energy detection performs rather poorly but is among the most popular techniques due to its low complexity. Moreover, it does not require extra information about the PU's signal and its statistics.

There are two important parameters to measure the performance of a spectrum sensing scheme: the probability of detection ($P_d$) and the probability of false alarm ($P_f$). The parameter $P_d$ is the probability that a SU correctly senses the existence of a PU in the frequency range of interest and the parameter $P_f$ is the probability that the SU mistakenly decides that the spectrum is occupied by a PU while it is not. The probability $1-P_d$ is proportional to the interference induced from the SU on the PU which must be as low as possible and the probability $1-P_f$ is proportional to the throughput of the SU. It is thus desirable to maximize $P_d$ and to minimize $P_f$.

Spectrum sensing can be performed as either distributed or cooperative [17]. In the distributed SS (DSS), each cognitive radio senses the spectrum continuously and then makes the decision on the availability of the spectrum individually. Although the distributed SS is simple and does not require additional overhead but sometimes it fails to make a correct decision especially when some SUs are shadowed or when the channel undergoes multipath fading and time dispersion [18]. To combat this problem, SUs can work in a cooperative manner so that the decisions made by them are combined with a predefined fusion rule [19, 20]. In this case, final sensing decision is used by all the SUs participant in the sensing process [21-25]. This approach not only resolves the problem of shadowed SUs but also improves the total probability of false alarm for a given total probability of detection and consequently improves the maximum throughput achievable by each SU.

Cooperative spectrum sensing (CSS) requires two preliminary steps. The first step is the formation of the cooperation cluster so that the members of the cluster are in each other's transmission range and that they can communicate their individual decisions properly with the minimum required bandwidth and error. The second required step is the allocation of the overhead bandwidth for each SU to broadcast their decisions to the other members of the cluster. Decisions passed between SUs can be either soft or hard [23].

When hard decisions are exchanged among SUs, each SU spends at least one bit of its available throughput to send its initial decision to the other SUs. A SU must wait until it receives decisions made by the other SUs to combine them with its own initial decision. Although the overhead associated with the communication of the initial decisions and the waiting time is rather negligible for small cluster sizes, it becomes more important as the cluster size increases. In fact, our study shows that there is an optimum cluster size which results in the maximum effective throughput.

In this paper, we study the effect of cooperation cluster size on the throughput of SUs in CSS and find the optimum cluster size.

The rest of paper is organized as follows. In Section II, the system model is presented. In Section III, the effective throughput is calculated as a function of cooperation cluster size. Numerical results are presented in Section IV and finally the paper is concluded in Section V.

## II. SENSING MODEL

Assume a simple cognitive radio system including 1 PU and several SUs, as shown in Fig. 1. Also, assume that $N$ of these SUs are in each other's transmission range. Each node can be either stationary or mobile. In a distributed scenario, each SU senses the spectrum continuously to detect whether the PU's spectrum is occupied or not. Without loss of generality, we assume that the SUs use energy detection to sense the spectrum and that each SU chooses its decision threshold based on the SNR of its own received signal. An energy detector makes a decision between two hypotheses $H_0$ and $H_1$ as follows

$$r_k = \begin{cases} n_k & H_0 \\ h s_k + n_k & H_1 \end{cases} \quad (1)$$

where $s_k$ and $n_k$ are the signal and the projection of the additive white Gaussian noise (AWGN) into the signal space at time index $k$, respectively, and $h$ is the channel gain which is assumed to be known at the receiver. Random variables $n_k$ and $h$ follow the complex Gaussian and Rayleigh distributions, respectively. The energy detector of the $i$th SU chooses the hypothesis $H_1$, i.e., senses the channel as occupied if the input of its decision device exceeds a predefined threshold as

$$\sum_{k=1}^{m_i} \left| r_k^i \right|^2 > \lambda_i \quad (2)$$

where $m_i$ and $\lambda_i$ are the sensing time and the threshold of the $i$th SU, respectively.

In cooperative spectrum sensing, since a SU does not use the spectrum when the other SUs are still sensing the spectrum, all SUs start sharing the free spectrum simultaneously and therefore equal sensing time for all members of the cluster provides the best performance. We thus assume $m = m_i$, $i = 1, 2, ..., N$.

The distribution of $\sum_{k=1}^{m_i} \left| r_k^i \right|^2$ for $H_0$ and $H_1$ scenarios in (1) is centralized and non-centralized Chi-square, respectively [8]. Therefore, the probabilities of detection and false alarm for each SU are calculated by the following equations, respectively [8].

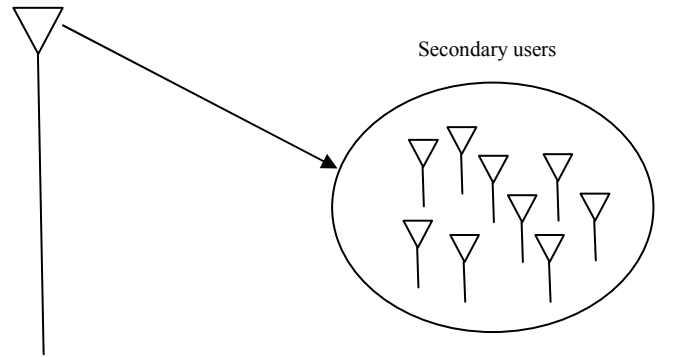

Fig. 1. A typical cognitive radio system with one PU and several SUs, all in the radiation range of the PU.

$$P_d^{\,i} = Q_m\left(\sqrt{2m\gamma_i}, \sqrt{\lambda_i}\right), \quad (3)$$

$$P_f^{\,i} = 1 - \Gamma\left(\frac{\lambda_i}{2}, m\right), \quad (4)$$

where $Q_m$ and $\Gamma$ are Marcum Q function and incomplete gamma function, respectively, and $\gamma_i$ is the SNR on the link between the PU and the $i$th SU (note that this is different from the SNR on the links between different SU's). The latter SNR affects the throughput of the SUs. It however, does not affect the sensing performance. The next step in the cooperative spectrum sensing is *sensing fusion*, where decisions made by the SUs are combined with some predefined fusion rule. In this paper, we consider the OR-rule and the AND-rule for sensing fusion. The total probabilities of detection and false alarm are calculated for each rule as follows:

$$P_d^{OR} = 1 - \prod_{n=1}^{N}\left(1 - P_d^{\,n}\right), \quad (5)$$

$$P_f^{OR} = 1 - \prod_{n=1}^{N}\left(1 - P_f^{\,n}\right), \quad (6)$$

$$P_d^{AND} = \prod_{n=1}^{N} P_d^{\,n}, \quad (7)$$

$$P_f^{AND} = \prod_{n=1}^{N} P_f^{\,n}. \quad (8)$$

In the following, we use the notation $P_d^{tot}$ to denote both $P_d^{OR}$ and $P_d^{AND}$. Similarly, $P_f^{tot}$ is used for both $P_f^{OR}$ and $P_f^{AND}$. The specific assignment is clear from the context, and depends on whether the fusion rule is the OR-rule or the AND-rule

In the spectrum sensing scenario, $1 - P_d^{tot}$ is proportional to the interference induced from each SU on the PU, and thus it must be as small as possible. From (3), (5), and (7), we can see that to have a larger $P_d^{tot}$, the threshold values $\lambda_i$ have to be decreased. This consequently leads to a larger $P_f^{tot}$. (In the limit, $P_d^{tot} = 1$ results in $P_f^{tot} = 1$, which corresponds to zero throughput.) There is thus always a trade-off between interference induced on the PU and the throughput of the SUs. In this work, our goal is to keep $P_d^{tot}$ above a certain threshold $1 - \varepsilon$, where $\varepsilon$ corresponds to the maximum interference tolerable by the PU. Before the $i$th SU starts the sensing, it must first calculate the threshold $\lambda_i$. Having $\varepsilon$, and assuming equal probability of the detection for the members of a cluster, we can calculate $P_d = P_d^{\,n}$ for each SU using (5) or (7). Then using (3), the threshold $\lambda_i$ for each SU is calculated.

III. THROUGHPUT VERSUS CLUSTER SIZE

Each SU spends $m$ symbols for spectrum sensing and then waits for the duration of at least $N$ symbols, one to send its decision to the other members of the cluster and $N$-1 to receive the decision made by the other members. Consequently, the effective throughput of each cognitive radio is

$$R_i = \left(1 - \frac{m+N}{T_s}\right)\left[1 - P_f^{tot}\left(P_d^{tot}, N, \gamma_1, \ldots, \gamma_N\right)\right]C_i, \quad (9)$$

where $C_i$ is the capacity of the $i$th SU, and $T_s$ is the total number of symbols available for sensing, communication among SUs, and transmission by SUs in one operational period of the SUs. We refer to $T_s$ as the *sensing period*. In (9), $P_f^{tot}$ is a function of both $N$ and $P_d^{tot}$, because for a given $P_d^{tot}$, the value of $N$ affects the threshold $\lambda_i$ through (3) and (5), or (3) and (7), which correspondingly affects $P_f^{tot}$ through (4) and (6), or (4) and (8). Since the value of $C_i$, which is a function of the SNR of SU-SU links, just scales the effective throughput, the optimum cluster size $N$ does not depend on it and therefore we normalize it to 1. On the other hand, since before the optimization of the cluster size, we do not exactly know the number and the identity of the cluster members including their $\gamma_i$ values, we approximate the right hand side of (9) by substituting all $\gamma_i$ values with $\bar{\gamma}$ which is the average SNR of the SUs that are in each other's transmission range. Consequently, (9) is normalized and approximated as

$$\bar{R}\left(P_d^{tot}, N, \bar{\gamma}\right) \approx \left(1 - \frac{m+N}{T_s}\right)\left[1 - P_f^{tot}\left(P_d^{tot}, N, \bar{\gamma}\right)\right]. \quad (10)$$

For given $\bar{\gamma}$ and $P_d^{tot}$, the optimum value of $N$ is calculated by

$$N_{opt} = \arg\max_{N}\left(1 - \frac{m+N}{T_s}\right)\left[1 - P_f^{tot}\left(P_d^{tot}, N, \bar{\gamma}\right)\right] \quad (11)$$

## IV. NUMERICAL RESULTS

In this Section, numerical results are presented for different values of $\bar{\gamma}$, $m$, and for both the AND-rule and the OR-rule. Figures 2 and 4 present the normalized throughput of SUs versus $N$ when respectively 5 percent and 20 percent of the sensing period is dedicated to spectrum sensing. Figures 3 and 5 present $P_f^{tot}$ corresponding to the normalized throughputs in Figures 2 and 4, respectively.

In all cases, for very low values of $\bar{\gamma}$, the throughput is monotonically increasing with the cluster size $N$. This is a consequence of the fact that for small $\bar{\gamma}$, the sensing result of individual SUs is not very reliable and thus cooperation with other SUs will improve the results. This improvement dominates the degradation of performance caused by the overhead required for the communication among SUs.

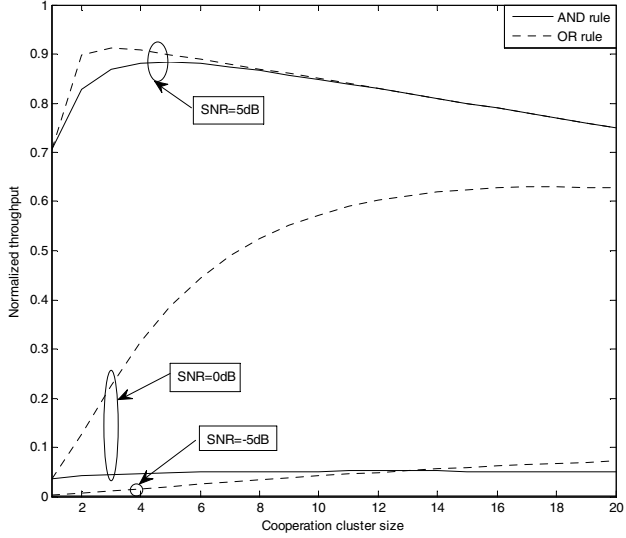

Fig. 2. Normalized achievable throughput versus the number of the cooperative nodes for $m = 0.05 T_s$.

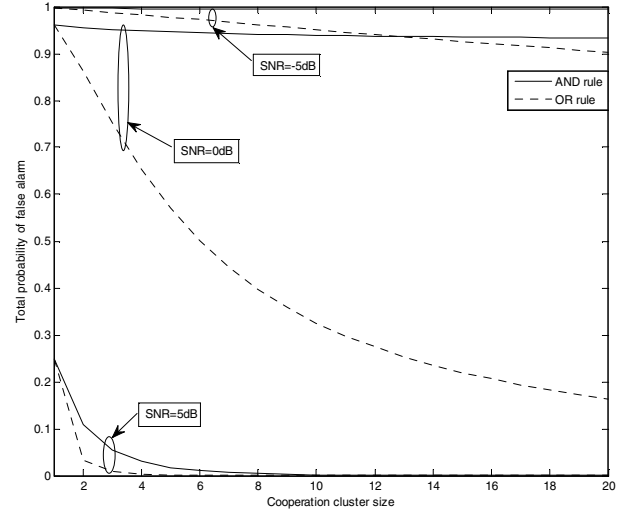

Fig. 3. Total probability of false alarm versus the number of the cooperative nodes for $m = 0.05 T_s$.

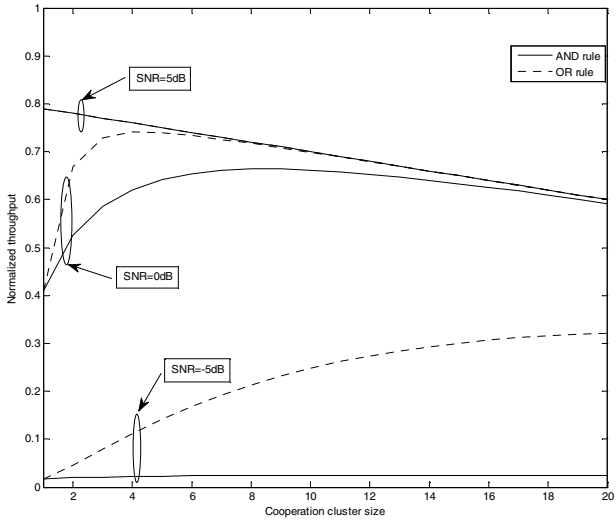

Fig. 4. Normalized achievable throughput versus the number of the cooperative nodes for $m = 0.2 T_s$.

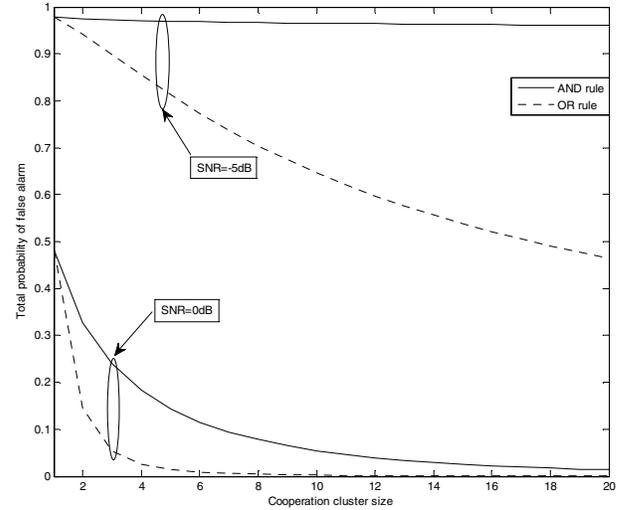

Fig. 5. Total probability of false alarm versus the number of the cooperative nodes $m = 0.2 T_s$.

This effect, according to Figures 2 and 4 is more prominent for the smaller values of $m$. From Fig. 2, we can see that when $\bar{\gamma} = 5\,dB$, the optimum value of $N$ is 2 if the OR-rule is used, and is 4 if the AND-rule is used. For the case of $m = 0.2T_s$ in Fig. 4, we can see that for $\bar{\gamma} = 5\,dB$, both the AND-rule and the OR-rule provide the same throughput and their curves monotonically decrease with the cluster size $N$. This implies that the optimum cluster size in this case is one. In other words, in this case, cooperation always decreases the effective throughput and distributed sensing presents a better result. For $\bar{\gamma} = 0\,dB$, optimum cluster sizes are 4 and 8 for the OR-rule and the AND-rule, respectively. In all cases, and when the cluster size is optimized, the OR-rule outperforms the AND-rule.

## V. CONCLUSION

In this paper, the effect of cooperation cluster size on the throughput of cooperative spectrum sensing is studied. It is demonstrated that there exists an optimal cluster size which maximizes the throughput of the secondary users. This optimal value is a function of the average SNR at the secondary users and the sensing time.

Various numerical results show that the optimum cluster size decreases with the increase in the average SNR. The results also indicate that for the AND-rule, the optimum cluster size is always larger than that of the OR-rule. Moreover, it is shown that when the cluster size is optimized, the OR-rule always provides a higher throughput compared to the AND-rule.